# Direct evidence for minority spin gap in the $Co_2MnSi$ Heusler alloy

Stéphane Andrieu*[1], Amina Neggache[1], Thomas Hauet[1], Thibaut Devolder[2], Ali Hallal[3], Mairbek Cschiev[3], Alexandre Bataille[4], Patrick Le Fèvre[5], Francois Bertran[5]

1 Institut Jean Lamour, UMR CNRS 7198 –Université de Lorraine- BP 70239, F-54506 Vandoeuvre cedex, France
2 Institut d'Electronique fondamentale – Université Paris Saclay
3 SPINTEC, CEA/CNRS, Grenoble
4 Laboratoire Leon Brillouin, CEA Saclay
5 Synchrotron SOLEIL, CNRS, St Aubin

This a draft version 0.000

Abstract : Half Metal Magnets are of great interest in the field of spintronics because of their potential full spin-polarization at the Fermi level and low magnetization damping. The high Curie temperature and predicted 0.7eV minority spin gap make the Heusler alloy $Co_2MnSi$ very promising for applications. We investigated the half-metallic magnetic character of this alloy using spin-resolved photoemission, *ab initio* calculation and ferromagnetic resonance. At the surface of $Co_2MnSi$, a gap in the minority spin channel is observed, leading to 100% spin polarization. However, this gap is 0.3 eV below the Fermi level and a minority spin state is observed at the Fermi level. We show that a minority spin gap at the Fermi energy can nevertheless be recovered either by changing the stoichiometry of the alloy or by covering the surface by Mn, MnSi or MgO. This results in extremely small damping coefficients reaching values as low as $7 \times 10^{-4}$.

Introduction : Although giant magneto-resistance was discovered more than 25 years ago, the development of electronics harnessing the spin of the electron -spintronics–is still very active due to a continuous flow of discoveries[Fer03]. Among them, the successive evidences of spin accumulation, spin transfer torque effects, their dependence on the crystal symmetries in magnetic tunnel junctions (MTJs), and more recently the multiple uses of spin-orbit coupling have exalted new research areas, including spinorbitronics. Spintronic devices rely on thin or ultra-thin layers of magnetic materials. In most spintronics studies, standard transition metal elements like Fe, Ni and Co (and alloys thereof) are conveniently used; however, alternative materials with superior electronic properties are desirable and are thus actively investigated. In particular, a class of materials called Half Metal Magnet (HMM) offers exciting properties both for theory and applications[DeG83]. By definition, HMMs have no minority spin states at the Fermi level ($E_F$): the material is thus a metal for majority spins and an insulator for minority spins[Gal07][Gra11]. Besides the interest of this full spin-polarization for transport properties, extremely low magnetic damping are also expected. Indeed, because of the minority spin gap, the magnons cannot find the spin-flipping electronic transitions at the Fermi level that systematically degrade the magnon lifetime in standard metallic magnets [Buz09][Liu09]. The expected combined low damping and perfect spin polarization of HMM would make them the perfect material for next generation spin transfer torque devices.

Such HMM properties were first predicted in magnetic oxides such as $Fe_3O_4$[Kat08], $CrO_2$[Sch86][Kam87], or $LaSrMnO_3$ (LSMO)[Pic97][Par98]. Yet, disappointing properties were found, including, for instance, small magnetizations at remanence or a too low Curie temperature[Fer03]. Other magnetic materials with high $T_c$, especially some ternary alloys belonging to the Heusler alloys families, were then explored theoretically[Gra11]. These materials were produced as thin film in the mid-1990s. The HMM property was claimed for many of them in order to explain transport properties, despite any direct evidence of a spin gap. The shortcoming of this approach was illustrated in the emblematic case of NiMnSb: while a low damping coefficient was measured, consistent with a HMM behavior in the bulk material[Pig10], the spin polarization measured using photoemission did not exceed 50% at the surface [Bon88] and felt to zero when covered by MgO[Sic05].

Lately, there was renewed interest after large tunnel magnetoresistance was obtained in $Co_2MnSi$/MgO-based MTJs [Sak06] [Ish09] [Liu12]. The predicted high spin polarization [Mei12] was observed very recently (93%) by spin-resolved photoemission experiments using 21.2eV photon energy[Jou14]. This has to be however confirmed for other photon energies to investigate some possible resonant surface state[Lou80]. Moreover, from a device perspective, it is essential to assess whether this property persists when covering the surface with a thin insulator such as MgO and whether the material has also a low magnetic damping. Here, we thus study the electronic and magnetic properties of $Co_2MnSi$ by combining spin resolved photoemission using various photon energies and ferromagnetic resonance spectroscopy. For the bare surface, we observe a total spin gap (100% spin polarization) for the minority spin band; however, this gap is located below the Fermi energy ($E_F$). At $E_F$, the spin polarization is reduced by the presence of a minority spin band observed in a very narrow photon energy range that was not explored in ref.[Jou14].Using stoichiometryengineering following theory [Wei12] and magneto-transport results [Liu12], we show that this potentially upsetting minority spin band can be shifted up in energy above $E_F$; it can even be fully suppressed by relevant surface hybridization when covered by Mn or MnSi, and, more interestingly, by MgO. This HMM character of the MgO covered surface, together with the observed extremely low magnetic damping make $Co_2MnSi$ one of the best material for next generation spintronic devices.

**Results:** In order to test the HMM character of $Co_2MnSi$ (noted CMS here), we first performed *ab initio* calculations considering the bulk Heusler structure (fig.1). This motivated the growth of thin CMS films by Molecular Beam Epitaxy (MBE). We aimed atsingle crystalline layers in order to take advantage of the orbital selectivity of the photoemission process by tuning the linear polarization of the photons.We have worked with MgO(001) substrates, whose lattice constant is close to CMS (001), and which also enablesthe production offully epitaxialMgO-based MTJs.

The electronic properties were analyzed by using Spin-Resolved PhotoEmissionSpectroscopy (SR-PES) using the SOLEIL synchrotron source,varying both the photon energy and polarization, to determine the symmetry of the states in the band structure [Joh97]. Electrons detection was performed along the sample normal, but alsooff-normal to explore the whole Brillouin Zone (see the supplementary information). As SR-PES is a surface sensitive technique, the films are grown in an MBE connected to the SR-PES analysis chamber[And14]. The Co, Mn and Si materials were co-evaporated onto the MgO (001) substrate held at a temperature around 800K. Reflection High Energy Electron Diffractionis used during the growth process. It showedthe expected lattice for the CMS surface structure. The chemical order was evidenced by new ½ streaks,observed along the (110) azimuth (see the supplementary informationand[Neg14]). Standard X-Ray diffraction analysis confirms the cubic structure typical of Heusler alloys. Finally, the films were heated in-situup to 1000Kjust after the growth [Ort13].

**Spin polarization of the uncoated surface:** As shown in fig.1, a spin gap of about 0.7eV is predicted in the calculated Density Of States (DOS) for bulk CMS (with no surface included in the calculation). Majority and minority spin PES spectra at room temperature for the bare CMS are shown in fig.2a. The 100% spin polarization is confirmed experimentally since the minority spin spectrum reaches zero for a binding energyaround -0.3eV. The spin gap is thus observed below $E_F$, while *ab initio* calculations performed considering bulk CMS predicted it to span upon $E_F$. It does not extend up to $E_F$ because of a minority spin contribution at $E_F$ (transition noted A in fig.2a). It should be noted that this contribution is not observed in DOS calculations in fig.1. Besides, there is also a majority spincontribution around -0.4eV (transition B in fig.2) which also does not appear in the bulk calculations. The A contribution was not observed in [Jou14], and a strong spin polarization was therefore found at $E_F$. This was due to the used photon energy (21.2eV) and probably also to a lower energy resolution (0.4eV) than the one available here (0.15eV). Indeed, we observed that these A and B transitions are onlydetected in a very narrow photon energy range from 35 to 40eV (see supplementary information).detected in a very narrow photon energy range from 35 to 40eV (see supplementary information). Any measurement out of this photon energy range thus misleadingly leads to strong positive spin polarization at $E_F$.

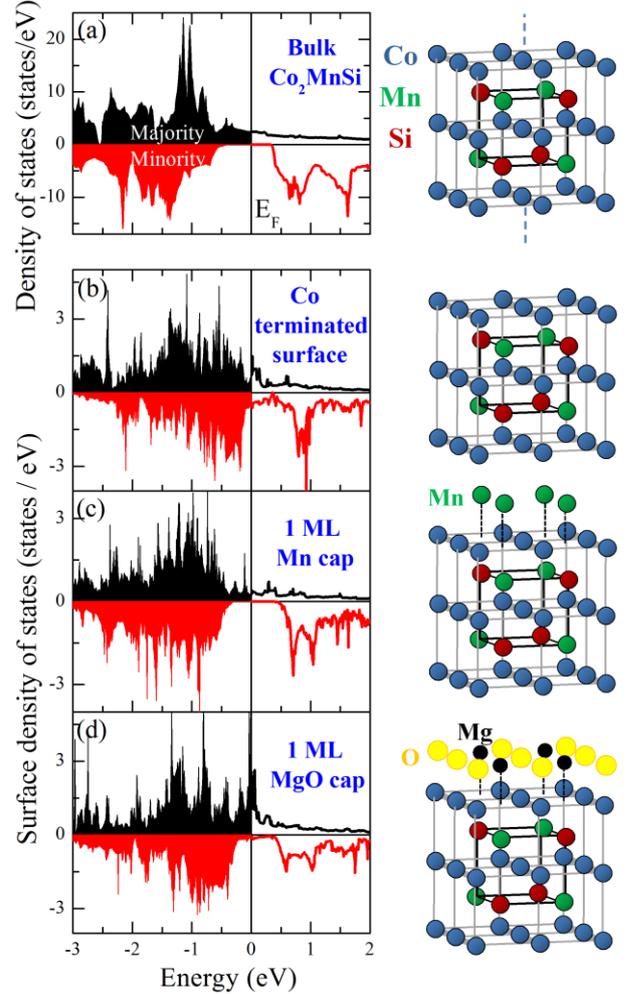

*Figure 1: (a) full density of states for both spins calculated for the bulk (with no surface) and surface contribution to the density of states for (b) a Co-terminated surface (c) 1 Mn atomic plane on top, and (d) 1 MgO atomic plane on top. The structures are shown on the right.*

To shed light on the origin of these A and B transitions, we have rotated the photon polarization in order to determine the symmetry of the states involved. The PES spectra obtained for *p* and *s* polarizations are compared in Fig.2. The *p* polarization exciteselectronic states with Δ1 and Δ5 symmetries, while the *s* polarization only excites Δ5 states. The loss of intensity, for the A and B transitions, when going from *p* to *s* polarization demonstrates that they are both of Δ1 character. Interestingly, when the use of *s* polarization suppresses the A and B transitions, the PES recovers consistency with the DOS calculated in the bulk: the spin gap now straddles the Fermi level down to -0.4eV binding energy, in very good agreement with calculations (fig.1). This is a strong indication that A and B contributions are linked to the surface. A majority spin surfaceresonant state was indeed predicted by Braun and co-workers[Bra15] using

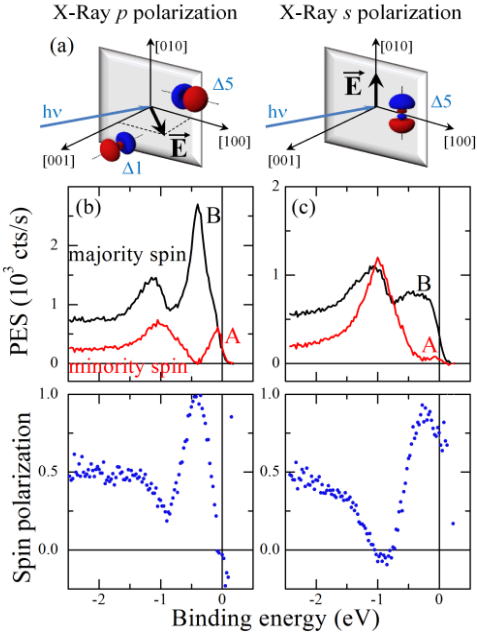

*Figure 2: schematic of the excited states depending on the light polarization (a) and spin-resolved and PES spectra using the p (b) and s (c) polarizations of the photons performed on the same CMS sample. The two peaks named A and B observed using the p polarization are strongly attenuated using s polarization, meaning that the initial state symmetry is Δ1. The spin gap is thus observed to extend up to $E_F$.*

*ab initio* calculations that can account for the B transition. However, it is calculated at 0.4eV above $E_F$ whereas the B transition is here observed below $E_F$. It seems to be also observed in a recent work using 6 eV photon energy, with the same polarization dependence [Fet15]. Finally, since those states are likely to be detrimental to magneto-transport properties, let us design strategies to mitigate their impact.

**Adjusting the spin polarization at $E_F$ :** Ishikawa et al[Ish09] studied magneto-transport in CMS/MgO/CMS MTJs and found a surprising result : the maximum of tunnel magnetoresistance was not observed for the exact CMS stoichiometry, but for an excess of Mn. To understand this behavior, we performed a SR-PES analysis on several samples with different Mn contents. We have studied both a simple excess of Mn and a partial replacement of Co by Mn, with qualitatively similar conclusions. PES spectra obtained for 1, 1.1 and 1.2 relative concentration of Mn in $Co_2Mn_xSi$ are plotted in fig.3. As Ishikawa et al. did[Ish09], we heated the CMS layers only up to 870K. At this temperature, the chemical ordering in the unit cell is incomplete, and correlatively, the spin polarization reaches 80% instead of the 100% that can be obtained after annealing above 1000K.

Increasing the Mn content shifts the maximum of the spin polarization peak towards $E_F$ which explains the conclusions drawn from magneto-transport. In fact, this shift of the spin gap can be understood from hand waving arguments. Indeed, replacing Co by Mn reduces the number of valence electrons available to fill the DOS. In a rigid band structure picture, this electron depletion moves the Fermi level closer to the spin gap as the Mn content is increased (fig.3). This also means that the minority spin contribution at $E_F$ formerly observed on samples with the exact stoichiometry is pushed towards empty states leading to high spin polarization at $E_F$ when enriching with Mn. Our experiments thus clearly explain the TMR increase when increasing the Mn content reported in ref.[Ish09].

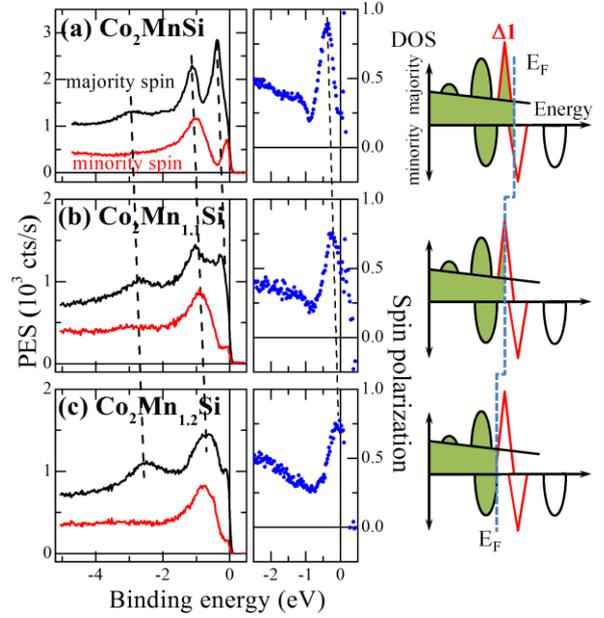

*Figure 3 : SR-PES spectra a 300K using p polarization and hν=37eV for a series of CMS layer with increasing Mn content. As shown in the schemes, the Fermi energy is moving towards the spin polarization maximum when increasing the Mn content.*

**Effect of surface coverage :** As we have seen in fig.1, the A and B contributions are not reproduced by calculations of the bulk DOS. One may assume that the A and B states result from chemical disorder in the bulk [Pic04]. In this case, capping the CMS would not change that situation, in stark contradiction with our results. A more plausible explanation is that the A and B transitions involve surface states. Several techniques are available to test this. First, we looked at the dispersion of these states along the (001) direction by varying the photon energy from 20 to 110 eV to scan the wavevector component perpendicular to the surface, the lack of dispersion of this component being a necessary condition for a surface state. Unfortunately, these transitions were only observed in range of photon energy from 35 to 40 eV, i.e. too narrow to assess the state dispersion (see supplementary information). This narrow photon energy range for excitation argues in favor of a resonant photoemission process already observed on surface states [Lou80] [Kev87]. If this is the case, it means that the states involved in the A and B transition have a strong *d* character. To conclude on the physical mechanism determining the surface spin polarization, we decided to cap the CMS with various overlayers. This follows the arguments developed by Galanakis et al [Gal02] who showed by *ab initio* calculations performed on $Co_2MnGe$ that the spin gap may be strongly affected by the surface termination. To verify whether this applies in our case, we calculated the DOS with different types of overlayers(fig.1). Whereas the spin gap is not obtained at $E_F$ with a bare CMS surface, in good agreement with ref.[Miu11], it is predicted to be retrieved when terminated with a Mn monolayer. We thus deposited a Mn monolayer (ML) on top of a CMS layer. We observed that the A and B contributions are strongly decreased, while the spin polarization strongly increased at $E_F$ (fig.4a) and c)). We also tested MnSi termination and observed the same effect. But more interestingly, as predicted by calculation (fig.1), a thin MgO termination is also efficient to retrieve the minority spin gap. Indeed, the measured spin polarization at $E_F$ increases up to around +70% (from 60 to 80% from one sample to the other) by adding MgO. This is larger than in previous works using also SR-PES, but with very low photon energy[Fet13][Fet15]. Moreover, an even larger spin polarization at $E_F$(90%) was obtained by using an Mn-enriched $Co_{1.8}Mn_{1.2}Si$ layer covered by MgO (Fig.4b) and d)). The possibility to tune the spin-polarization at $E_F$ both with CMS stoichiometry and MgO covering is a very promising result. It contrasts with the NiMnSb Half-Heusler alloy case, for which the spin polarization is destroyed when MgO is deposited on the surface [Sic06]. Altogether, these experiments are a strong indication that A and B contributions are probably similar bands originating from the surface and separate in energy by exchange splitting of the electronic states due to the ferromagnetic behavior. Although the agreement between the calculated DOS and measured PES is not quantitative, the observed enhancement of spin polarization at $E_f$ due to MgO and Mn coverage (as compared to free CMS) is qualitatively explained.

CMS layers and after the capping are shown in (c) –Mn cap- and (d) -MgO cap-. As the spin polarization is close to zero at $E_F$ for both CMS uncoated surfaces, it is strongly increased when covering with Mn or MgO. The reason is the strong attenuation of A and B contributions in spin–resolved spectra.

**Gilbert damping:** Gilbert damping can be viewed as a measure of the decay rate of the population of low energy magnons. In magnetic metals, magnons decay mostly when they are annihilated by collisions with electrons whose spin is flipped in that process [Kam76]. Since the energy of the magnons that matter for magnetization dynamics lies in the 10 µeV range, the transitions involve only electronic states very close to the Fermi energy. These transitions do not exist in HMM, such that their damping should be comparable to that of insulators, i.e. lower than that of Fe (alpha=0.0019), which is the lowest damping among metallic magnets [Dev13]. We thus measured the dynamic magnetic properties of Au covered CMS samples using Vector Network Analyzer FerroMagnetic Resonance (VNAFMR) [Bil08] in both in-plane and out-of-plane applied fields. The FMR frequency-field relation (fig.5a) provides the effective magnetization. The latter appeared to depend strongly on the temperature at which the CMS was post-growth annealed. The largest magnetizations (1.25T) were obtained for samples heated at 1000K, where chemical order was the most complete.

On these samples, the FMR lines were particularly narrow, with raw linewidths (fig.5b) already smaller than the best ever reported in iron. An estimate of the Gilbert damping can be done by fitting the evolution of the linewidth with the resonance frequency in perpendicular applied field. We found Gilbert damping values of $10^{-3}$ on several stoichiometric samples, and a record value of $7.10^{-4}$ observed once on a $Co_{19}Mn_{11}Si_{10}$ sample. Such values are extremely low for a conductive material, and are consistent with a HMM behavior at least in the bulk of CMS.

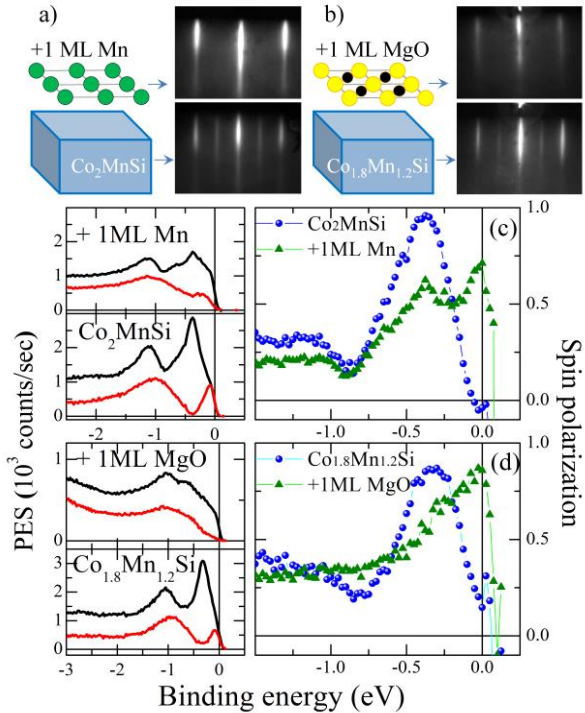

*Figure 4: influence of Mn and MgO coverage on CMS electronic properties. In (a) and (b) are shown the electron diffraction patterns for the starting CMS layers and when covered by Mn or MgO. The PES and spin polarization spectra for the starting*

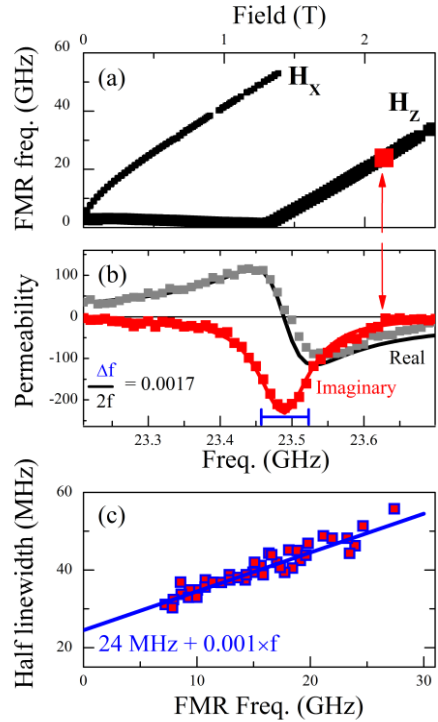

*Figure 5: FMR properties of a 20 nm thick CMS film. (a) Frequency versus field plots for in-plane (Hx) and out-of-plane (Hz) applied fields. (b) Real and imaginary parts of the exeprimental (symbols) and modeled (lines) permeability for an out-of-plane field of 2.1 T. (c) Frequency linewidth versus FMR frequency in Hz field. The line is a linear fit.*

**Discussion:** In summary, we experimentally established that the CMS Heusleralloy is a half-metal magnet in its bulk. The half metallic character is altered at the surface by the presence of surface states of Δ1 character, close to the Fermi levelin both minority and majority spin channels. Enriching the CMS alloy with Mn lowers its Fermi energy below the surface states. The Fermi energy is then in the spin gap and the material becomes half metallic at its surface too. This clarifies the dependence of tunnel magnetoresistance on the alloy stoichiometry. Moreover, we also proved that these statesare strongly linked to the surface termination of CMS, and that they can be almost suppressed when covering CMS with Mn, MnSi and MgO. The HMM character comes with extremely low values of Gilbert damping, which reaches values never observed before on conducting materials. All of these results make the CMS/MgO system an excellent candidate for the study of spin related phenomena and the related applications.

**SUPPLEMENTARY INFORMATIONS**

Calculation Method: Calculations are performed using Vienna ab initio simulation package (VASP) [Kre93, Kre96] with generalized gradient approximation [Wan91] and projected augmented wave pseudopotentials [Kre99, Blö94]. We used the kinetic energy cutoff of 550 eV and a Monkhorst-Pack k-point grid of $11 \times 11 \times 1$ for the surface calculations and $9 \times 9 \times 7$ for the bulk. Initially the structures were relaxed until the force acting on each atom falls below 1 meV/°A.

Sample preparation: The sample preparation and SR-PES measurements were performed at the CASSIOPEE beamline of the SOLEIL synchrotron radiation facility. The CMS samples are grown in a MBE set-up connected to the SR-PES chamber. We used 2 e-gun sources for Co and Si and a Knudsen cell for Mn. The growth rates were accurately controlled during the growth using quartz microbalances. The epitaxial process was controlled by using electron surface diffraction (RHEED) and the chemical quality of the films was controlled by Auger spectroscopy available in the MBEchamber (supfig.1).The CMS surface is very stable since no change on the RHEED patterns was observed after 24h PES measurements; the Auger analysis showed very small surface contamination by C and perhaps O (mixed with theMn-LMM Auger transitions). This small contamination yet decreases the intensity of the resonant surface state after long measurements (but not the other PES transitions).The ½ streaks on RHEED pattern along [11] azimuth is due to the chemical order [Neg14]. The CMS surface covering was done by Mn, MnSi or MgOevaporations,calibrated using a quartz microbalance. It should be noted that the ½ streaks discussed above are not seen anymorewhen Mn or MgO are deposited, which is expected according to Mn and MgO lattices. The MgO deposition is performed using an e-gun source. The MgO growth rate was checked using RHEED oscillations.

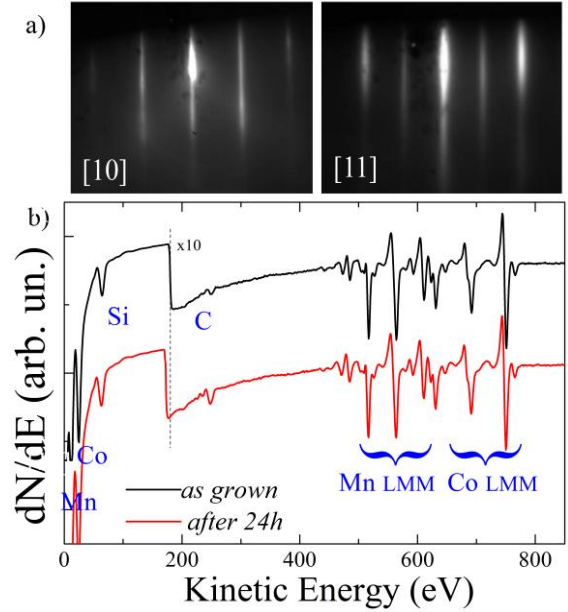

*Sup-fig.1: (a) Typical RHEED patterns on CMS and (b)Auger spectra just after the growth and after PES measurements (24h).*

Photoemission experiments: Photoemission was carried out in a spin-resolved photoemission set-up using a Mott detector (with a Sherman coefficient equal to 0.12) with a 150 meV energy resolution. The spin detector is equipped with 4 channels that allowed us to measure spectra for both in-plane and out-of-plane spin components. The normal of the sample was oriented along the axis of the detector. The detector aperture is fixed at it maximum equal to ±8°. This corresponds to 50% of the Brillouin Zone (BZ) at 37eV photon energy at normal incidence. To explore the whole BZ, we rotated the sample off-normal by 8° from the detector axis. PES spectra performed at 37 and 43 eV for 0° and 8° are shown in supfig.2. Similar transitions are observed for both measurement angles,meaning that the whole BZ is explored. Only the surface resonant states (A and B contributions) decreased at 8° off-normal meaningthat these states are located alongΓ.

To eliminate instrumental asymmetry in spin-resolved measurements, two distinct PES spectra were recorded with opposite magnetizations. As the coercive field of our CMS films was less than 20 Oe, the applied magnetic field was set to 200 Oe to saturate the magnetization prior to PES measurements. We systematically observed a zero out-of-plane spin-polarization, confirming that the magnetization of our CMS films was in-plane.The symmetry of the initial states involved in the transitions observed on SR-PES spectra was determined by measuring successively with s or p polarizations of the incoming photons. As the photon beam was at 45° from the normal to the sample, *s* polarized photons excite only occupied Δ5 states, whereas *p* polarized photons excite both occupied Δ1 and Δ5 states [Bon12]. The SR-PES analysis was performed using incoming photon energy ranging from 25 to 90 eV. The surface resonant states

# ARTICLES

are observed in a very narrow range of photon energy as shown in supfig.3. This was confirmed by Angle-Resolved Photoemission experiments varying the photon energy from 20 to 110eV.

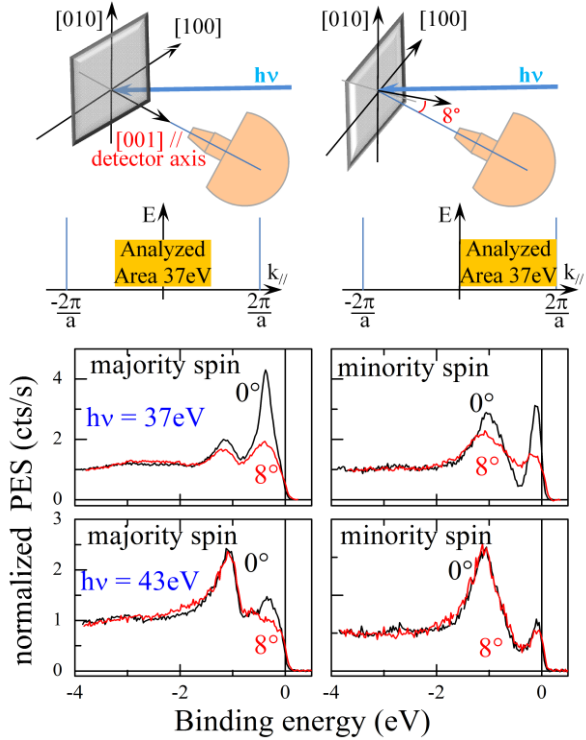

Sup-fig.2: top, schematics showing the Brillouin Zone area explored at 37eV depending on the angle between [001] and the detector axis. Bottom, PES spectra dependence with this angle (0 or 8°) for 37 and 43 eV photon energies.

The PES measurements were performed both at 300K or 80K. We do not observe any significant influence of the sample temperature on the PES results. Around 30 samples were grown and measured during 4 runs of 1 week each.

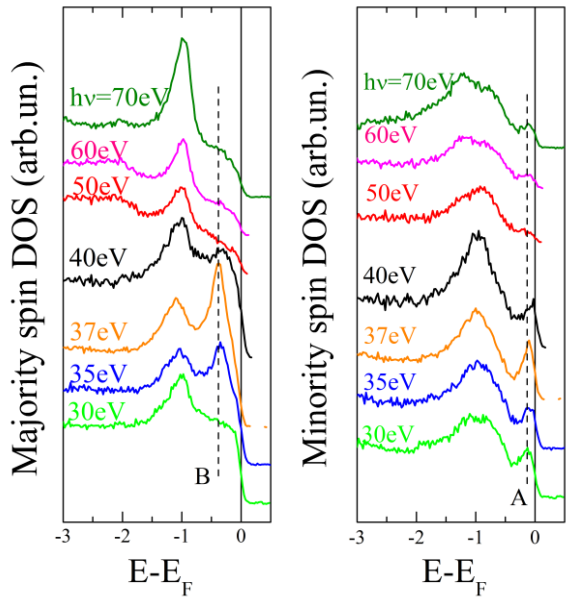

Sup-fig.3: PES spectra dependence with the photon energy, for both majority and minority spins. The B contribution is observed for photon energies between 35 and 40 eV.

# ARTICLES

**Author contributions**
S. A. conceived the study. S.A., A.N. and T.H. fabricated the samples. P.L.F. and F.B. designed the PES instrument. S.A., A.N., T.H, A.B.T.D., P.L.F. and F.B performed the PES measurements. T.D. performed the FMR measurements. M.C. and A.H. calculated the band structures and the density of states. All authors analyzed and interpreted the data. S.A. prepared the manuscript. All authors commented on the manuscript.

**Acknowlegmements**
We would like to thank A.D. Kent from NYU for his critical reading of the manuscript, and C. Fontaine from Rennes University (France) for fruitful discussion on the PES results. This work was supported by the Region Lorraine.

**Competingfinancialinterests**
The authors declare that they have nocompeting financial interests.